# A Privacy-Protecting Framework of Autonomous Contact Tracing for SARS-CoV-2 and Beyond

Shamiul Alam[1], Md Shafayat Hossain[2], and Ahmedullah Aziz[1*]

[1]Dept. of Electrical Eng. & Computer Sci., University of Tennessee, Knoxville, TN, 37996, USA
[2]Dept. of Electrical Engineering, Princeton University, Princeton, NJ, 08544, USA
[*]Corresponding Author's Email: aziz@utk.edu.

*Abstract*— Controlling the spread of infectious diseases, such as the ongoing SARS-CoV-2 pandemic, is one of the most challenging problems for human civilization. The world is more populous and connected than ever before, and therefore, the rate of contagion for such diseases often becomes stupendous. The development and distribution of testing kits cannot keep up with the demand, making it impossible to test everyone. The next best option is to identify and isolate the people who come in close contact with an infected person. However, this apparently simple process, commonly known as - contact tracing, suffers from two major pitfalls: the requirement of a large amount of manpower to track the infected individuals manually and the breach in privacy and security while automating the process. Here, we propose a Bluetooth based contact tracing hardware with anonymous IDs to solve both the drawbacks of the existing approaches. The hardware will be a wearable device that every user can carry conveniently. This device will measure the distance between two users and exchange the IDs anonymously in the case of a close encounter. The anonymous IDs stored in the device of any newly infected individual will be used to trace the risky contacts and the status of the IDs will be updated consequently by authorized personnel. To demonstrate the concept, we simulate the working procedure and highlight the effectiveness of our technique to curb the spread of any contagious disease.

## 1. Introduction

Traditionally, during the outbreak of contagious diseases, infected people get calls from public health workers regarding their contacts, and those contacts are then asked to quarantine/self-isolate. This process is known as contact tracing. It has been used for decades to control the spread of communicable diseases, such as tuberculosis, vaccine-preventable infections (such as measles), sexually transmitted infections (including HIV), Ebola, bacterial infections, and novel infections, e.g. H1N1, SARS-CoV, and SARS-CoV-2 (COVID-19) [1]. More recently, contact tracing is undergoing new developments, particularly in the context of COVID-19. To reduce the spread of COVID-19, a few countries have been employing different contact tracing approaches [2]. For instance, South Korea has been maintaining a public database including information about the travel routes of the infected people [2]. Israel is using the cell-phone data to track the movement of infected individuals [3]. In Taiwan, medical institutions were given access to the travel histories of the patients [4]. Singapore has developed a mobile app to track the Bluetooth data on contact [5]. All these approaches suffer from serious privacy concerns because the government, medical institutes, and (in some cases) general people are given access to very sensitive data. There have been reports of software-based approaches [6], [7] to address the privacy concerns, but none are safe from potential cyber-security threats. For instance, Apple and Google have been collaborating on a contact tracing application that will utilize cryptography to ensure anonymity [8]. However, this approach demands that every individual should carry a smartphone. It will be infeasible to deploy such a system on the national/global level because of the immense overhead, especially in the under-developed countries. On the other hand, in the beginning of COVID-19 pandemic in 2020, the Centers for Disease Control and Prevention (CDC) of the USA have expressed the need to employ 100,000 contact tracers to track the spread of COVID-19 [9]. The Association of State and Territorial Health called on the US Congress to allocate 7.6 billion USD, solely for contact tracing in the USA [10]. Clearly, there exists a dire need to establish a more efficient, reliable and automated contact tracing system to effectively curb the relentless spread of COVID-19 and any future pandemic.

In this work, we propose a hardware-based, privacy-protecting contact tracing approach based on a wearable contact tracing device (CTD). From a system-level perspective, the key idea for this device is straightforward. Every individual will wear/carry a CTD anytime they go out of their homes. The prime task of a CTD will be to automatically detect the presence of another CTD within a critical distance (e.g., 6 feet for COVID-19). Once such proximity is detected, the pair of CTDs will exchange and store hardware tags/IDs that are unique to themselves. The hardware ID of a CTD will only be known to its owner and will not be linked with any other personal data or location information. Therefore, our proposed technique offers a potential gateway towards a contact tracing which will ensure the protection of the users' private information. Although the ID of one's CTD is not linked to any personal information about the patient, the privacy protection will still rely on the authorized personnel who is responsible for protecting the patient's medical records (such as COVID positive/negative records). Therefore, there is no possibility of any breach of private information of the users and the protection of the data will only rely on the ethics of the authorized medical personnel. The CTDs will utilize Bluetooth low energy (BLE) [11] to seamlessly communicate with each other, similar to the smartphone-based approach. However, our proposed CTDs will be built as an application-specific, mass-producible, and ubiquitous tool for contact tracing.

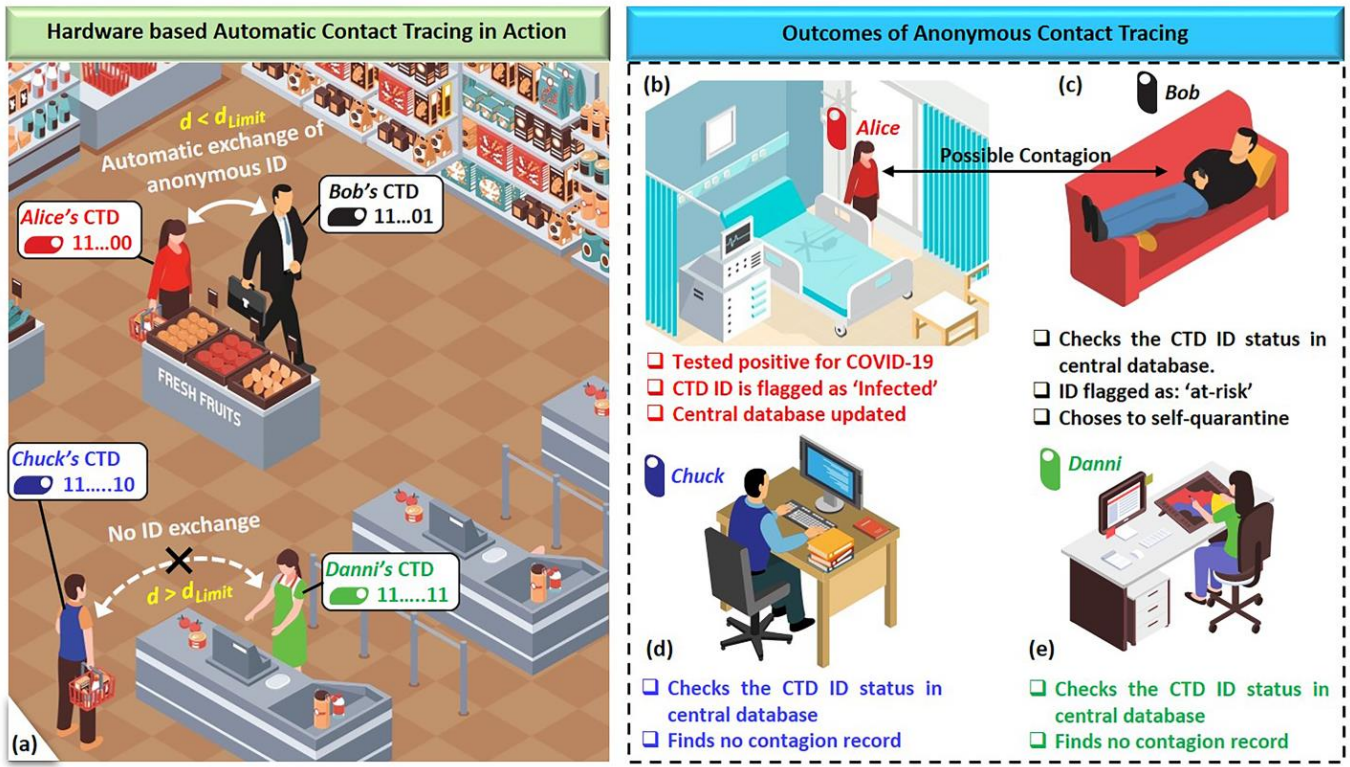

**Fig. 1:** Working principle of the proposed contact tracing approach. (**a**) Detection of a CTD and exchange of anonymous IDs in the case of a close contact ($d < d_{Limit}$). (**b**) Alice, one of the CTD users, is tested positive for COVID-19, and the IDs saved in the NVM of her CTD are flagged accordingly. (**c**) Due to the close contact with Alice, the recently reported "infected" individual, Bob's ID is flagged to "at-risk" and self-isolation is suggested. (**d**) & (**e**) Chuck and Danni, respectively, did not come in close contact with any "infected" or "at-risk" person, hence they had not received any quarantine instructions.

The major contributions of this work are:
- We propose a hardware-based autonomous contact tracing framework that will protect the privacy of the users.
- To demonstrate the contact tracing utilizing our technique, we present a human mobility simulation and show the effectiveness of our proposed technique to control the spread of any infectious disease.

## 2. Privacy-Protecting Contact Tracing Device

We illustrate our proposed contact tracing approach in Fig. 1 using a simple example. Here, to explain the concept, we consider four users wearing/carrying the proposed CTD. The BLE module in the device (also see Fig. 3) detects the nearby CTDs and exchanges the anonymous IDs if conditions are fulfilled [IDs are shared for the case of Alice and Bob in Fig. 1(a)]. The measured distance is compared with the previously set critical limit of the distance ($d_{Limit}$), and for every contact with distance lower than $d_{Limit}$, the IDs are stored in the built-in non-volatile memory (NVM) block (Fig. 3) if the ID doesn't already exist. The working principle of the proposed CTD based contact tracing approach is shown using a flow chart in Fig. 2. Whenever any CTD user is tested positive for COVID-19, the authorized personnel will flag the ID of the person as 'infected' in the central database [Fig. 1(b)]. Subsequently, the status of the IDs saved in that CTD are marked as 'at-risk' due to possible exposure [Bob's ID is marked as 'at-risk' due to close contact with Alice, shown in Fig. 1(c)]. The CTD users can use the web portal of the central database using any edge device (e.g. computer, cell phone, etc.) to check the status of their CTD ID. In severely resource-constrained localities, a central system can be set up to announce the 'at-risk' IDs in print or digital format. The key benefit is that the contact tracing efforts will not be limited by the number of smart devices. A single internet-connected device can be used to check the status of thousands of IDs, leading to an exponential expansion of the scope of contact tracing.

Figure 3 shows the central components of the proposed contact tracing device. The BLE module is responsible for detecting critical proximity and exchanging hardware IDs. Although our proposed technique is compatible with other wireless technologies, we use BLE because BLE is the most suitable technology for distance measurements because of its high scan rate, very low power consumption, widespread commercial availability, and better signal geometry [12], [13]. The received signal strength indicator (RSSI) of the BLE modules can be utilized to estimate the distance between the transmitting and the receiving devices using the well-established path loss model [14]. This model correlates the signal strength, physical distance, and environmental factors through the following equation:

$$RSSI = -10\, n\, \log_{10}(d) + C, \qquad (1)$$

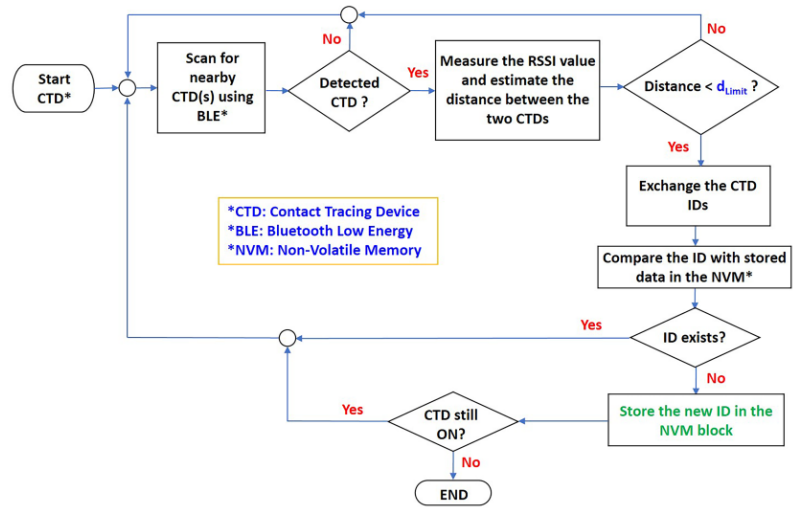

**Fig. 2:** Flow chart of the proposed contact tracing approach.

where, n is the environment-dependent path loss exponent, d is the distance between the transmitting and receiving devices, and C is a constant that accounts for system losses. This model can also be used to sense physical barriers between two individuals (e.g. wall, window, etc.) and thereby avoid marking such proximity events as risky. A CTD will be equipped with three memory devices: (i) Read only memory (ROM) – to store the anonymous ID of the CTD, (ii) Random access memory (RAM) – for on-the-fly processing of RSSI data received by the BLE module, and (iii) Non-volatile memory– to store the IDs of other CTDs that come in close contact (see Fig. 3). A timestamp corresponding to the proximity occurrences can also be stored. The entire CTD system shown in Fig. 3, assembled in a wearable fashion, will be powered by a rechargeable battery. A universal serial bus (USB) interface will allow connection with external edge devices to read out the stored IDs. A custom made low-power central processing unit (CPU) will manage the coordination and communication among all these modules.

## 3. Human Mobility Simulation Results

To demonstrate the contact tracing based on our CTD, we present a simulated scenario of ten CTD users in an area. In our simulation, we categorize the total area into four regions: (i) work zone, (ii) community zone, (iii) residential zone, and (iv) others. The human mobility in these regions is modeled via dividing the whole day (Day-1) into three segments: (i) work hours (8 am – 5 pm), (ii) community hours (5 pm – 8 pm), and (iii) residential hours (8 pm – 8 am). We define a few terminologies to indicate the health condition of a person. A person is marked 'infected' if he/she officially tests positive for COVID-19. People who came in close contact ($d < d_{Limit}$) with 'infected' individuals (or their contacts) will be marked - 'at-risk'. Finally, the individuals who never came in close contact with 'infected' or 'at-risk' people will be marked - 'not-at-risk'. Fig. 4 shows the simulated result for an arbitrary weekday. The location of a person at any given time is also randomly chosen from within the allocated area. We assume that the day starts with two 'infected' individuals, chosen at random. As the day progresses, these two people (red markers - Fig. 4) come in close contact with some previously healthy individuals (green markers - Fig. 4) and lead to the risk of contagion. These 'at-risk' individuals (blue markers - Fig. 4) can act as secondary sources of infection and hence their contacts will also be marked 'at-risk'. Fig. 4 presents selected time snaps from the simulated human mobility throughout a day to illustrate these scenarios.

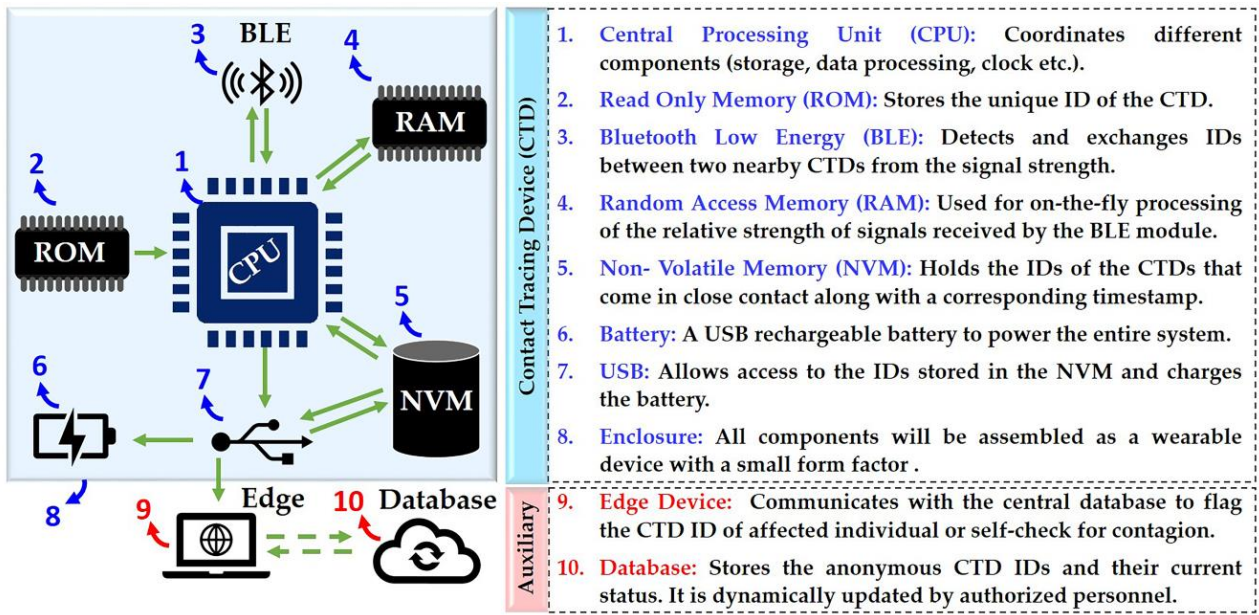

**Fig. 3:** Components of the proposed contact tracing device.

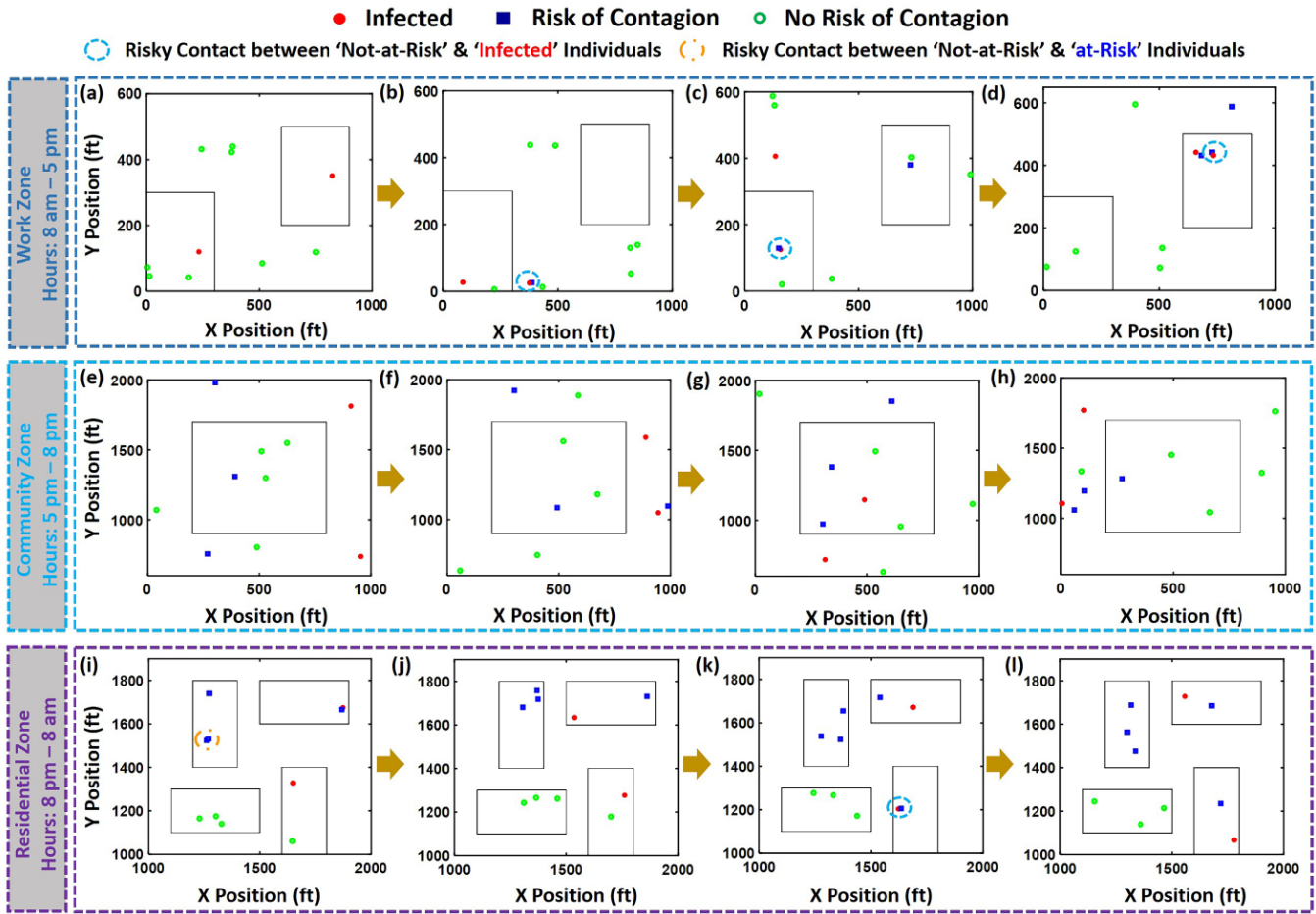

**Fig. 4:** Simulated results to show the CTD based contact tracing approach for 10 people in an area. Random travel routes are generated for the people in different zones of the area at different time spans of the day. Black lines (boundary of the boxes) represent the walls of the buildings in the considered area. Throughout the entire day (Day-1) we show human interactions within (**a**) - (**d**) work zone, (**e**) - (**h**) community zone and (**i**) - (**l**) residential zone. Incidences of close contacts ($d < d_{Limit}$) between a 'not-at-risk' and 'infected' individuals are marked in (**b**), (**c**), (**d**) & (**k**) and a close encounter between a 'not-at-risk' and 'at-risk' people is marked in (**i**).

The outcome of the contact tracing process after Day-1 is summarized in Fig. 5(a). It shows that, five people who were initially 'not-at-risk', had been involved in risky contacts on Day-1 (marked with blue and yellow colors). Four of them had come in close contact with 'infected' people (marked with blue) and one of them had come in contact with an 'at-risk' individual (marked with yellow). When these people access the central database to check their status, they will receive different notifications and recommendations as per the status of their CTDs. Once an individual is marked 'at-risk', he/she will be expected to voluntarily reach out to medical professionals to get tested and to trace the contacts stored in their own CTDs. This voluntary conduct will be crucial to extend the scope of contact tracing beyond the primary source of contagion. Considering the possibility of human error or intentional misconduct, only authorized medical professionals may be allowed to add 'at-risk' IDs to the central database. Depending on the need, hardware security measures can be incorporated to prevent manipulation of the stored data in these CTDs.

Finally, for any contact tracing approach to be effective in controlling the spread of infectious diseases, it is crucial to follow the recommendations based on the contact tracing result. To examine the effectiveness of our proposed contact tracing approach, we have simulated for the second day (Day-2) using the abovementioned contact tracing results obtained for Day-1. We consider the following two cases: (i) only the 'infected' individuals are self-quarantined [Fig. 5(b)] and (ii) all the "infected" individuals and three (out of five) 'at-risk' individuals are self-quarantined [Fig. 5(c)]. For the (i), as shown in the contact tracing result in Fig. 5(b), after Day-2, all the people will be 'at-risk' due to the close contacts with the 'at-risk' individuals of Day-1. However, in the case (ii), where three out of five 'at-risk' people (after Day-1) quarantined themselves, no new people have become 'at-risk'. Figs. 5 (b & c), therefore, substantiate that the spread of any communicable disease can be controlled by effective contact tracing and by following the recommendations based on the contact tracing.

Here, we present a human mobility simulation of ten people in an area to demonstrate the contact tracing based on our proposed technique. The computational complexity of this simulation will depend on the number of CTD users in the simulation and also on the number of movements considered for each CTD user. The computational complexity will exponentially increase with the increase in number of users and number of movements considered for each user. For example, in our simulation for 10 users, there are total 45 ($10_{C_2}$) possible combinations of contacts for each of the movements when the distance will be measured and the condition for exchanging IDs will be checked. Now, if we consider a scenario with 100 or 1000 users in an area, there will be total 4950 ($100_{C_2}$) or 499500 ($1000_{C_2}$) possible contacts for each movement respectively. That means, there will be an exponential increase in the number of total possible contacts with the increase in the number of users for each movement. Now, for a realistic simulation considering a large number of users in an area and with a significant number of movements of each user in a day, the computation will be very complex and might require a high-performance computing cluster.

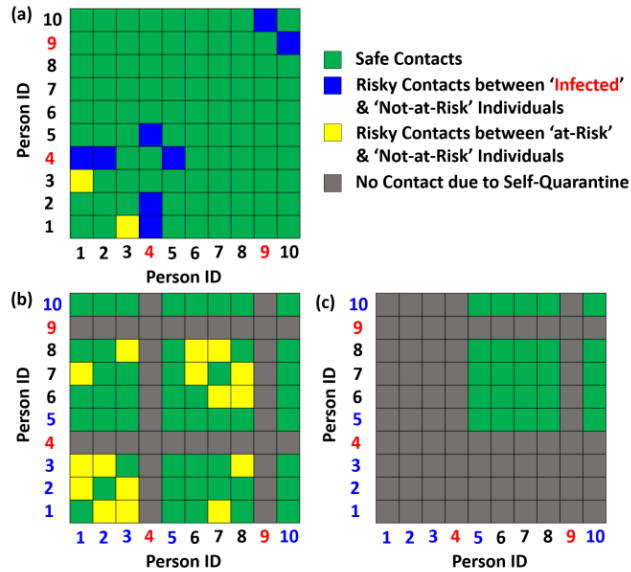

**Fig. 5:** Outcome and effectiveness of the anonymous contact tracing using our proposed technique. (**a**) Representation of the close contacts based on the saved IDs in the non-volatile memories of the CTDs on Day-1. We consider two scenarios to illustrate the importance of contact tracing and following the corresponding guidelines. (**b**) Case-I: we simulate the movement of ten people throughout a day (Day-1) and assume that only 'infected' individuals isolate themselves the next day (Day-2). We examine and present the incidences of widespread risky contacts occurring throughout Day-2 (identified by the simulated CTDs). (**c**) Case-II: we again simulate the movement of ten people throughout Day-1 and assume that all the 'infected' individuals and three (out of five) 'at-risk' individuals isolate themselves on Day-2. Because of the isolation of 'infected' individuals and their contacts, no further risk of contagion is seen after Day 2.

## 4. Ethical Concerns of Contact Tracing

Traditionally, the appropriate use of the contact tracing process has always relied on public trust and goodwill. Such a reliance entails the concerns regarding ethics, morality, guarantees of equity of access and treatment, effective oversight of patient's data, transparent protocols of contact tracing, etc. [15]. Our scheme offers a potential gateway towards a contact tracing without heavy reliance on public ethics and morality. First, the CTD IDs will be anonymous, which will protect the privacy of its user and will make it impossible to trace the health information back to the patient. That being said, it still relies on the authorized medical personnel being responsible for protecting the patient's medical records. Therefore, the requirement of careful oversight and protection of the data will only rely on the ethics of the authorized medical personnel.

Furthermore, our proposed device and the database will run on a simple algorithm that can be explained to the users. There is no Blackbox-type of machine learning in use for our scheme. Thus, the transparency of the protocol will be maintained.

Lastly, the assurance of equity of access and treatment is of paramount importance. Our CTD-based approach is inherently inclusive because the CTD will be provided to everyone by the government agencies, which will share the same technology. Therefore, everyone will be treated equally, which is different from the app-based approach where there are obvious advantages from the use of expensive smartphones and cellular networks.

## 5. Conclusions

In summary, we propose a privacy-preserving wearable contact tracing device to control the spread of COVID-19 and any other communicable diseases. Our next plan is to build and implement a prototype of this device in the near future. Note, in real scenario, the distance measurement will be affected by different environmental factors and system losses (such as interference in the surrounding area, fading of the transmitted signal etc.) which can result into false positives or negatives. Therefore, our proposed CTD needs to be implemented in hardware with cautions so that the non-ideal effects due to the environmental factors and system losses can be minimized. We firmly believe that if we can encourage every individual of a locality to wear our CTD, the contact tracing for COVID-19 and possible future pandemics will be easier, efficient, less expensive, and more privacy-protecting than other existing contact tracing approaches. In our simulation, we have also underscored the necessity to follow the recommendations based on the contact tracing system. Our simulation clearly highlighted the possibility of the aggressive spread of a contagious disease, in the absence of a coordinated contact tracing and isolation procedure. We also recognize that the effectiveness of our proposed system will rely on the responsible conduct and awareness of the individuals, similar to any other non-authoritarian approach.

**Author Contributions:** A.A. and M.S.H. conceived the primary idea. S.A. and M.S.H. performed the simulations and data analysis. All authors discussed the results and took part in writing the manuscript. A.A. directed the overall project.

**Funding:** This research received no external funding.

**Conflicts of Interest:** The authors declare no conflict of interest.

**References:**
[1] K. T. D. Eames and M. J. Keeling, "Contact tracing and disease control," *Proc. R. Soc. B Biol. Sci.*, 2003, doi: 10.1098/rspb.2003.2554.
[2] I. A. Hamilton, "11 countries are now using people's phones to track the coronavirus pandemic, and it heralds a massive increase in surveillance - We Are The Mighty." [Online]. Available: https://www.wearethemighty.com/mighty-survival/countries-tracking-coronavirus-patients-phones/?rebelltitem=1#rebelltitem1. [Accessed: 18-Jun-2021].
[3] J. Tidy, "Coronavirus: Israel enables emergency spy powers - BBC News." [Online]. Available: https://www.bbc.com/news/technology-51930681. [Accessed: 18-Jun-2021].
[4] C. J. Wang, C. Y. Ng, and R. H. Brook, "Response to COVID-19 in Taiwan: Big Data Analytics, New Technology, and Proactive Testing," *JAMA - Journal of the American Medical Association*. 2020, doi: 10.1001/jama.2020.3151.
[5] H. Cho, D. Ippolito, and Y. W. Yu, "Contact Tracing Mobile Apps for COVID-19: Privacy Considerations and Related Trade-offs," Mar. 2020.
[6] T. Altuwaiyan, M. Hadian, and X. Liang, "EPIC: Efficient Privacy-Preserving Contact Tracing for Infection Detection," in *IEEE International Conference on Communications*, 2018, vol. 2018-May, doi: 10.1109/ICC.2018.8422886.
[7] L. Reichert, S. Brack, and B. Scheuermann, "Privacy-Preserving Contact Tracing of COVID-19 Patients," *IACR Cryptol*, 2020.
[8] Apple-Google, "Privacy-Preserving Contact Tracing - Apple and Google," *COVID19-contractting*, 2020. .
[9] P. Sullivan, "CDC director: US has about 28,000 contact tracers, needs 100,000 | TheHill." [Online]. Available: https://thehill.com/policy/healthcare/504098-cdc-director-us-has-about-28000-contact-tracers-out-of-100000-needed?__twitter_impression=true. [Accessed: 18-Jun-2021].
[10] P. Sullivan, "Urgency mounts for a contact tracing army | TheHill." [Online]. Available: https://thehill.com/homenews/coronavirus-report/501416-urgency-mounts-for-a-contact-tracing-army. [Accessed: 18-Jun-2021].
[11] P. Malekzadeh, A. Mohammadi, M. Barbulescu, and K. N. Plataniotis, "STUPEFY: Set-Valued Box Particle Filtering for Bluetooth Low Energy-Based Indoor Localization," *IEEE Signal Process. Lett.*, 2019, doi: 10.1109/LSP.2019.2945402.
[12] J. Luo and H. Gao, "Deep Belief Networks for Fingerprinting Indoor Localization Using Ultrawideband Technology," *Int. J. Distrib. Sens. Networks*, 2016, doi: 10.1155/2016/5840916.
[13] S. Sadowski and P. Spachos, "RSSI-Based Indoor Localization with the Internet of Things," *IEEE Access*, 2018, doi: 10.1109/ACCESS.2018.2843325.
[14] P. Kumar, L. Reddy, and S. Varma, "Distance measurement and error estimation scheme for RSSI based localization in wireless sensor networks," in *5th International Conference on Wireless Communication and Sensor Networks, WCSN-2009*, 2009, pp. 80–83, doi: 10.1109/WCSN.2009.5434802.
[15] L. Ferretti, C. Wymant, M. Kendall, L. Zhao, A. Nurtay, L. Abeler-Dörner, M. Parker, D. Bonsall, and C. Fraser, "Quantifying SARS-CoV-2 transmission suggests epidemic control with digital contact tracing," *Science (80-. ).*, 2020, doi: 10.1126/science.abb6936.